\documentstyle[12pt,epsf]{article}
\begin{document}

\bigskip\bigskip
\begin{center}

{\large \bf Transverse momentum, factorization\\ 
and the HERMES experiment}

\end{center}
\vspace{12pt}

\begin{center}
 {\bf Elvio Di Salvo\\}

 {Dipartimento di Fisica and I.N.F.N. - Sez. Genova, Via Dodecaneso, 33 \\-
 16146 Genova, Italy\\}
\end{center}

\vspace{10pt}
\begin{center} {\large \bf Abstract}

I present some results about transverse momentum dependent distribution and 
fragmentation functions. Firstly I illustrate a simple model, with predictive 
power about the energy behavior, for T-odd, chiral odd functions. Moreover I 
propose a slight modification in extracting transversity from HERMES data, so 
as to apply correctly factorization. Lastly I suggest a method for determining 
the quark transverse polarization in an unpolarized or spinless hadron. 
\end{center}

\vspace{10pt}

\centerline{PACS numbers: 13.85.Qk, 13.88.+e}

\newpage

\section{Introduction}

It is well-known that the transversity function[1-4] is quite difficult to 
determine experimentally[5-9], as well as all chiral odd functions. The most 
promising methods are based on the azimuthal asymmetries, which are sensitive 
to transverse momentum dependent (t.m.d.) distribution and fragmentation 
functions. A big contribution in this sense has been given by the HERMES 
experiment\cite{her}, whose data analysis is strongly based on the T-odd 
functions\cite{coll} and on the factorization assumption\cite{coll,qi1,tmp}, 
essential for extracting tranversity. In this talk I revise that 
analysis, proposing improvements which allow to apply correctly 
factorization. I also present a simple model for T-odd functions, predicting 
their behaviors as functions of the energy scale $Q$. 

After a short review of the azimuthal asymmetries (sect. 2), I consider (sect. 
3) the single spin asymmetry: I define the cross sections and asymmetries
measured in the HERMES experiments, moreover I introduce the correlation 
matrix, dedicating a particular attention to the T-odd functions, for which I 
have elaborated a simple model. In sect. 4 I revise the HERMES data analysis, 
proposing a modified weight function for extracting transversity and the 
distribution function $h_1^{\perp}$, which is related to the quark transverse 
polarization in an unpolarized proton. In sect. 5 I suggest an alternative 
method for extracting the T-odd, chiral odd fragmentation function for the 
pion. Lastly I give a short summary in sect. 6.

\section{Azimuthal asymmetries}

The t.m.d. functions were introduced at first 
by Ralston and Soper\cite{rs}. The importance of such functions has been 
pointed out in the last years\cite{tmp,mt,bdr}, since when people started to 
plan measurements of azimuthal asymmetries[16-19]. Indeed, such 
functions contain nontrivial information on the internal structure of the 
nucleon\cite{mt}. In particular, T-odd functions allow, in principle, to 
measure transversity in a semi-inclusive deep inelastic scattering (SIDIS)
experiment\cite{coll,ja01,ef}. In order to extract t.m.d. functions, one 
usually considers the weighted asymmetries[17-19], which will be 
defined in the following sections. Here I shall be concerned with asymmetries 
relative to two types of experiments:

a) SIDIS\cite{her,smc,he1}:
\begin{equation}
\ell {\vec p} (p^{\uparrow}) \to \ell' \pi X, \label{r1}
\end{equation}
$\ell$ denoting a charged lepton;

b) Electron-positron annihilation into two jets, observing a final 
pion\cite{ja01,ef}:
\begin{equation}
e^+ e^- \to \pi X; \label{r2}
\end{equation}
In particular reaction (\ref{r2}) allows to extract a T-odd fragmentation
function\cite{coll}, to be used in the HERMES data analysis of reaction 
(\ref{r1}), in order to get transversity\cite{ja01,ef}. This procedure 
is based on factorization[11-13], which has not been proven for t.m.d. 
functions. However this property is generally assumed, provided the energy 
scale $Q$ is not too large, otherwise the Sudakov damping\cite{bo}
highly suppresses the asymmetry and deeply modifies its dependence of on t.m.d. 
functions.

\section{Spin asymmetries at HERMES}
\subsection{General formulae}
The HERMES experiments concern reaction (\ref{r1}). In particular, the 
experiment which has been realized\cite{her} employes a longitudinally 
polarized proton target, while for the next future the use of a transversely 
polarized target has been planned\cite{he1}. In one-photon approximation, the 
essential part of the reaction consists of
\begin{equation}
\gamma^* {\vec p} (p^{\uparrow}) \to \pi X. \label{r4}
\end{equation}
But the proton polarization - be it longitudinal or transverse with respect to 
the incident lepton beam - has always a longitudinal and a tranverse component 
with respect to the virtual photon momentum, which is the most relevant 
direction in this approximation. Since 
the Hamiltonian of a quantum system is linear with respect to spin, the effects 
on these two components of the polarization may be studied independently of 
each other. This 
is why transversity can be inferred also from a longitudinally polarized target.

I consider reaction (\ref{r4}) with transverse 
polarization, for which I define the single spin azimuthal asymmetry
\begin{equation}
A = \frac{d\sigma_{\uparrow}-d\sigma_{\downarrow}}
{d\sigma_{\uparrow}+d\sigma_{\downarrow}}.\label{asy0}
\end{equation}
Here $d\sigma_{{\uparrow}(\downarrow)}$ refers to cross sections with opposite 
polarizations. Moreover 
\begin{eqnarray}
d\sigma_{\uparrow}-d\sigma_{\downarrow} &\propto& d\Gamma L^{\mu\nu}
H^a_{\mu\nu},\label{diff0}
\\
L_{\mu\nu} &=& k_{\mu}k'_{\nu} +k'_{\mu}k_{\nu}-g_{\mu\nu}k\cdot k',\label{lt}
\\
H^a_{\mu\nu} &=& \int d^2p_{\perp} Tr\left[\gamma_{\mu}\Phi_{\chi.o.}
(x,{\bf p}_{\perp})\gamma_{\nu}{\hat \Phi}_{\chi.o.}(z,{\bf P}_{\perp})\right].
\label{tens}
\end{eqnarray}
$d\Gamma$ is the phase space element. $k$ and $k'$ are the initial and final 
four-momentum of the lepton. $\Phi_{\chi.o.}$, $x$ and ${\bf p}_{\perp}$ are, 
respectively, the chiral odd component of the correlation matrix\cite{mt}, the
longitudinal fractional momentum and the transverse momentum of a quark inside 
the proton. ${\hat \Phi}_{\chi.o.}(z,{\bf P}_{\perp})$ is the chiral odd 
component of the correlation matrix of a quark fragmenting into a pion, whose 
longitudinal fractional momentum and transverse momentum are respectively $z$ 
and ${\bf P}_{\perp}$. This last is such that the transverse momentum of the 
pion with respect to the photon, {\it i. e.},
\begin{equation}
{\bf \Pi}_{\perp} = {\bf P}_{\perp}+z{\bf p}_{\perp},\label{cons}
\end{equation}
is kept fixed.  

\subsection{Parametrization of the T-odd correlation matrix}

I consider a frame where the proton is transversely polarized and has a large 
momentum, say ${\cal P}$. In this frame the chiral odd component of the 
correlation matrix can be parametrized as
\begin{equation}
\Phi_{\chi.o.} = \frac{1}{4}x{\cal P}\gamma_5\left\{[\rlap/S,\rlap/n_+]
\left(h_{1T} + h'_1\frac{p_1}{\mu}\right) +[\rlap/n_1,\rlap/n_+]h^{''}_1
\frac{p_2}{\mu}\right\}. \label{chir}
\end{equation}
Here $h_{1T}$ is the t.m.d. tranversity function, while
$h'_1$ and $h^{''}_1$ are two new functions, to be illustrated below. Moreover
\begin{eqnarray}
p_1 &=& {\bf p}_{\perp}\cdot {\bf S}\times {\bf n}, ~~~~ \ ~~~~ \  p_2 = 
{\bf p}\cdot{\bf S},
\\ 
\ \sqrt{2} n_+ &\equiv& (1, {\bf n}), ~~~~ \ ~~~~~ \ ~~~~ \  n_1 \equiv 
(0, {\bf S}\times {\bf n}).
\end{eqnarray}
${\bf S}$ and ${\bf n}$ are unit vectors respectively in the direction of the 
proton polarization and of the proton momentum. Lastly $\mu$ is an undetermined 
mass scale, which was set equal to the  the proton rest mass by various 
authors\cite{rs,mt,tm}; as I shall show in the next subsection, this is not 
the most suitable 
choice. Furthermore, in reaction (\ref{r4}), it is convenient to 
identify ${\cal P}$ with $Q/2x$, where $Q$ is the virtual photon momentum; this 
amounts to considering a Breit frame where the active  quark has an initial 
longitudinal momentum equal to half photon momentum, but in the opposite 
direction\cite{fey}. 

The second and third term of parametrization (\ref{chir}) are T-odd and give 
a nonvanishing contribution also in the case of an unpolarized proton, since 
they are even under the exchange $S$ $\to$ $-S$. I show that the sum of these 
two terms must be independent of $S$. To this end, I consider the probability
density $q_{\uparrow}$ = $q_{\uparrow}(S;x,{\bf p}_{\perp})$ for a quark to 
have a positive spin component along a given unit vector ${\bf s}$ not parallel
to {\bf n}. I distinguish a T-even and a T-odd component of this density:
\begin{equation}
q_{\uparrow}(S;x,{\bf p}_{\perp}) = q^{(e)}_{\uparrow}(S;x,{\bf p}_{\perp}) +
q^{(o)}_{\uparrow}(S;x,{\bf p}_{\perp}).\label{up}
\end{equation}
The combined action of parity inversion and time reversal yields
\begin{equation}
q_{\downarrow}(-S;x,{\bf p}_{\perp}) = q^{(e)}_{\uparrow}(S;x,{\bf p}_{\perp}) -
q^{(o)}_{\uparrow}(S;x,{\bf p}_{\perp}),\label{dw}
\end{equation}
where $q_{\downarrow}$ is the probability density for a quark to have a 
negative spin component along ${\bf s}$. Analogous relations can be written for 
$q_{\downarrow}(S;x,{\bf p}_{\perp})$. Then, in an unpolarized proton, the 
difference $q_{\uparrow}-q_{\downarrow}$ results in   
\begin{eqnarray}
\delta q_{\perp} &=&\frac{1}{2}\left\{q_{\uparrow}(S;x,{\bf p}_{\perp}) -
q_{\downarrow}(S;x,{\bf p}_{\perp}) - \left[q_{\uparrow}(-S;x,{\bf p}_{\perp}) -
q_{\downarrow}(-S;x,{\bf p}_{\perp})\right]\right\} \nonumber 
\\
&=& q^{(o)}_{\uparrow}(S;x,{\bf p}_{\perp})
-q^{(o)}_{\downarrow}(S;x,{\bf p}_{\perp}) = \delta q^{(o)}_{\perp}.
\end{eqnarray}
But the quark density matrix of an unpolarized proton is $S$-independent, and 
the same must be of the quark polarization $\delta q_{\perp}$ =
$\delta q^{(o)}_{\perp}$. To impose this condition in eq. (\ref{chir}), I set
\begin{equation}
h'_1 = -h^{''}_1 = h_1^{\perp}.
\end{equation}
Therefore the sum of the last two terms of eq. (\ref{chir}) can be written as
\begin{equation}
\Phi^T_{\chi.o.} = \frac{x{\cal P}}{4\mu}\gamma_5[\rlap/r_{\perp},\rlap/n_+]
h_1^{\perp}, \label{todd}
\end{equation}
where
\begin{equation}
r_{\perp} = p_1S - p_2n_1 \equiv (0,-p_2,p_1,0).
\end{equation}
Eq. (\ref{todd}) implies 
\begin{equation}
\delta q_{\perp} = -\frac{r_{\perp}\cdot s_0}{\mu}h_1^{\perp}, 
\label{polr}
\end{equation}
where $s_0\equiv (0,{\bf s})$. Eq. (\ref{polr}) exhibits the meaning of the 
function $h_1^{\perp}$, already introduced by Boer and Mulders\cite{bm}.

In the case of quark fragmenting into a pion, a  T-odd, chiral odd 
fragmentation function can be defined in quite a similar way, substituting 
${\cal P}$ by $z{\cal P}$ and ${\bf p}_{\perp}$ by ${\bf P}_{\perp}$. Such a 
function - to be denoted as ${\hat h}_1^{\perp}$ in the following - describes 
the Collins effect\cite{coll}.

\subsection{A model for T-odd functions}

A proton may be viewed as 
a bound state of the active quark with a set $X$ of spectator partons.
In order to take into account coherence effects, I project the bound 
state onto scattering states with a fixed third component of the total 
angular momentum with respect to the proton momentum, $J_z$, and with a spin 
component $s =\pm 1/2$ of the quark along the unit vector {\bf s} introduced 
in subsect. 3.2. For the sake of simplicity, I assume that $X$ has spin zero, 
moreover I choose a state with $J_z$ = 1/2. Then 
\begin{equation}
|J_z=1/2; s; X\rangle = \alpha |\rightarrow, L_z=0; s; X\rangle + \beta 
|\leftarrow, L_z=1; s; X\rangle.
\end{equation}
Here $\rightarrow(\leftarrow)$ and $L_z$ denote the components along 
${\bf n}$, respectively, of the quark spin and orbital angular momentum, while
$\alpha$ and $\beta$ are Clebsch-Gordan coefficients. 
Then the probability of finding a quark with $J_z$ = 1/2 and spin component $s$ 
along ${\bf s}$, in a longitudinally polarized proton with a positive helicity, 
is
\begin{eqnarray}
|\langle P,\Lambda =1/2|J_z=1/2; s; X\rangle|^2 &=& \alpha^2 
|\langle P,\Lambda =1/2|\rightarrow, L_z=0; s; X\rangle|^2 \nonumber
\\
&+& \beta^2 |\langle P,\Lambda =1/2|\leftarrow, L_z=1; s; X\rangle|^2 + I, 
\label{prob}
\end{eqnarray}
\begin{equation}
I = 2 \alpha \beta Re\left[\langle P,\Lambda =1/2|\rightarrow, L_z=0; s; 
X\rangle 
\langle (\leftarrow, L_z=1; s; X)|P,\Lambda =1/2 \rangle\right].\label{intf} 
\end{equation}
Expanding the amplitudes in partial waves yields
\begin{equation}
I = 2 \sum_{l,l'=0}^{\infty} Re\left[ie^{-i\phi} A_l B^*_{l'}\right]
P_l(cos\theta) P^1_{l'}(cos\theta). \label{inter}
\end{equation}
Here $A_l$ and $B_l$ are related to partial wave amplitudes; moreover
$\theta$ and $\phi$ are respectively the polar and the azimuthal angle of 
the quark momentum, assuming ${\bf n}$ as the polar axis and, as the azimuthal 
plane, the one through ${\bf n}$ and ${\bf s}$. In the Breit frame one has
\begin{equation}
P_l(cos\theta) \sim 1, ~~~~~~ \ ~~~~~ P^1_l(cos\theta) \sim 
\frac{|{\bf p}_{\perp}|}{x{\cal P}}.
\end{equation}
Then eq. (\ref{inter}) yields
\begin{equation}
I \sim \frac{|{\bf p}_{\perp}|}{x{\cal P}}\left(A cos\phi + B sin\phi\right),
\label{intff}
\end{equation}
where $A$ and $B$ are real functions made up with $A_l$ and $B_l$. Since ${\bf 
s}$ is an axial vector, parity conservation implies $A$ = 0. Therefore eqs. 
(\ref{prob}) and (\ref{intff}) imply that the interference term $I$ is T-odd
and that the final quark is polarized perpendicularly to the proton momentum 
and to the quark momentum, independent of the proton polarization. Comparing 
eq. (\ref{intff}) with eq. (\ref{polr}) yields 
\begin{equation}
\mu = x{\cal P}. \label{norm}
\end{equation}
A similar reasoning can be applied to the Collins effect, for which one has 
\begin{equation}
\mu = xz{\cal P}. \label{norm1}
\end{equation}
Eqs. (\ref{norm}) and (\ref{norm1}) predict 
that both the Collins effect and the quark tranverse polarization in an 
unpolarized (or spinless) hadron decrease as ${\cal P}^{-1}$, where, as 
illustrated in subsect. 3.2, ${\cal P}$ is related to the energy scale $Q$.

\section{Data Analysis in SIDIS}
\subsection{Extracting the transversity function from HERMES data}

Eqs. (\ref{diff0}), (\ref{tens})  and (\ref{chir}) imply that the numerator
of the SIDIS asymmetry (\ref{asy0}) is of the form
\begin{equation}
d\sigma_{\uparrow}-d\sigma_{\downarrow} \propto 
\int d^2p_{\perp} \left[\frac{P_1}{xz{\cal P}} h_{1T} + 
\frac{{\bf p}_{\perp}\cdot{\bf P}_{\perp}}{zx^2{\cal P}^2}
h_1^{\perp} \right]{\hat h}_1^{\perp}, \label{conv}
\end{equation}
assuming the constraint (\ref{cons}). Here 
\begin{equation}
P_1 = {\bf P}_{\perp}\cdot {\bf S}\times{\bf n}.
\end{equation}
In order to extract the transversity function, {\it i. e.}, $h_1$ = $\int 
d^2p_{\perp}h_{1T}$, I define the following weighted asymmetry:
\begin{equation}
\langle A_1\rangle = \frac{\sum_nd\sigma^{(n)}\Pi_1^{(n)}}{M_P
\sum_nd\sigma^{(n)}},
~~~~~~~ \ ~~~~~~ \Pi_1 = P_1+zp_1. \label{asy}
\end{equation}
Here $M_P$ is the proton rest mass and $d\sigma^{(n)}$ the differential cross 
section at a fixed transverse momentum, the sum running over the data. Eq. 
(\ref{conv}) implies
\begin{equation}
\sum_nd\sigma^{(n)}\Pi_1^{(n)} \propto h_1(x) {\hat h}_{1(1)}^{\perp}(z),
\ ~~~~~~ \ ~~~~~ \ 
{\hat h}_{1(1)}^{\perp}(z) = \int d^2P_{\perp} P_1^2{\hat h}_1^{\perp}.
\label{we}
\end{equation}
This allows to extract $h_1(x)$, provided the function 
${\hat h}_{1(1)}^{\perp}(z)$ is known. This function - which does not coincide 
with the Collins function\cite{coll,her} and from now on will be named the 
modified Collins function - has to be inferred from an independent experiment, 
for example from reaction (\ref{r2}), as I shall exhibit in the next section. 

\subsection{How to single out the function $h_1^{\perp}$}

$h_1^{\perp}$ can be determined by using the weighted asymmetry
\begin{equation}
\langle A'_1\rangle = \frac{\sum_n(d\sigma_{\uparrow}^{(n)}-
d\sigma_{\downarrow}^{(n)})(\Pi_1^{(n)})^2}{M^2_P\sum_nd\sigma^{(n)}}.
\label{asy1}
\end{equation}
Indeed, formula (\ref{conv}) yields
\begin{equation}
\langle A'_1\rangle \propto h_{1(1)}^{\perp}(x) {\hat h}_{1(1)}^{\perp}(z),
\label{we1}
\end{equation}
where $h_{1(1)}^{\perp}(x)$ is defined analogously to the second formula 
(\ref{we}), with ${\bf p}_{\perp}$ instead of ${\bf P}_{\perp}$. Also this 
function can be inferred only if 
the modified Collins function ${\hat h}_{1(1)}^{\perp}(z)$ is  known. 
Formula (\ref{conv}) implies that the weight functions ${\bf \Pi}^2_{\perp}$ 
and $({\bf \Pi}_{\perp}\cdot{\bf S})^2$ could be used instead of $\Pi_1^2$.

\section{Determining the modified Collins function}

The modified Collins function can be extracted from reaction (\ref{r2}). The 
helicity of the virtual, timelike photon is directed along the direction of the 
two initial lepton beams in the laboratory frame. Therefore the final pion 
distribution presents an azimuthal asymmetry around the jet direction, with 
respect to the plane $\Omega$ containing the jet and beam directions. In this 
case the asymmetry is proportional to the difference
\begin{equation}
d\sigma'_{\uparrow}-d\sigma'_{\downarrow} \propto d\Gamma' L'_{\mu\nu} Tr\left\{
\gamma^{\mu}{\hat \Phi}_{\chi.o.}\gamma^{\nu}\rho_{\chi.o.}\right\} \propto
{\Pi'_1} {\hat h}_1^{\perp}. 
\label{col39}
\end{equation}
Here $\Pi'_1$ = ${\bf \Pi}'_{\perp}\cdot {\bf n}_{\perp}$, where ${\bf 
\Pi}'_{\perp}$ is the transverse momentum of the pion and ${\bf n}_{\perp}$ 
the unit vector perpendicular to $\Omega$. Moreover $d\Gamma'$ is the phase 
space element and $L'_{\mu\nu}$ the symmetric leptonic tensor (\ref{lt}), made 
up with the four-momenta of the two initial leptons. Lastly $\rho_{\chi.o.}$ 
is the chiral odd component of the density matrix of the parton whose 
fragmentation is not observed.

I adopt the weighted asymmetry (\ref{asy}), substituting $\Pi_1$ with 
$\Pi'_1$. Owing to eq. (\ref{col39}), this asymmetry, which I call
$\langle A_1^c\rangle$, results in
\begin{equation}
\langle A_1^c\rangle \propto {\hat h}_{1(1)}(z).\label{coll2}
\end{equation}

The weight function I have suggested for SIDIS and for $e^+ e^-$ annihilation
differs from the one used in the data analysis of the HERMES experiment, 
{\it i. e.},
\begin{equation}
sin \phi = \frac{\Pi_1}{|{\bf \Pi}_{\perp}|},\label{weg}
\end{equation}
where $\phi$ is known as the Collins angle. This weight function, unlike those 
proposed in this talk, does not allow to factorize the integral in the first 
eq. (\ref{we}) into a function of the sole $x$ times a function of the sole 
$z$. This is due to the factor $|{\bf \Pi}_{\perp}|^{-1}$. For the same reason,
if (\ref{weg}) is used, the T-odd fragmentation function extracted from 
$e^+e^-$ annihilation data is different from the one which 
appears in the formula of SIDIS asymmetry. Therefore it is quite difficult to 
extract $h_1$ by means of this method. The modification proposed (see the 
second eq. (\ref{asy})) is essential for factorization.

I have neglected quark flavor throughout my talk. Including this quantum number 
would involve problems similar to those pointed out by E. Leader in
this conference.

\section{Summary}
Here I recall the main  results illustrated in the present talk.
\begin{itemize}
\item I have elaborated a simple, and yet sufficiently general, model for T-odd 
functions, which allows to predict the $Q$-dependence of such functions.
\item I have defined some weight functions, different from those used 
previously, and in particular a modification to the Collins function, which 
allows to apply correctly factorization in 
extracting chiral odd functions - and above all transversity - from data.
\item In particular I have proposed the weighted asymmetry $\langle A_1^{(c)}
\rangle$ (eq. (\ref{coll2})) for extracting the modified Collins function 
from reaction (\ref{r2}). Moreover I have suggested the weighted asymmetries 
$\langle A_1 \rangle$ (first eq. (\ref{asy})) and 
$\langle A'_1\rangle$ (eq. (\ref{we1})) for determining, respectively, the 
transversity and the function $h_1^{\perp}$ from reaction (\ref{r1}).
\item According to my model, $\langle A_1 \rangle$ and $\langle A_1^{(c)}
\rangle$ decrease as $Q^{-1}$, while $\langle A'_1\rangle$ decreases as 
$Q^{-2}$.
\end{itemize}

\end{document}